\documentclass[12pt,preprint]{aastex}

\usepackage{psfig}

\usepackage{emulateapj5}

\newcommand{\ic}{IC~443}
\newcommand{\ASCA}{{\sl ASCA}}
\newcommand{\Ginga}{{\sl Ginga}}
\newcommand{\Einstein}{{\sl Einstein}}
\newcommand{\sax}{{\sl BeppoSAX}}
\newcommand{\Chandra}{{\sl Chandra}}
\newcommand{\ROSAT}{{\sl ROSAT}}

\newcommand{\h}{$^{\rm h}$}
\newcommand{\m}{$^{\rm m}$}
\newcommand{\s}{$^{\rm s}$}
\newcommand{\ratio}{$R_{\mathrm{H/He}}$}

\slugcomment{submitted to ApJ 2001 November 21; accepted 2002 February 22}

\shorttitle{ASCA Observations of SNR IC~443}
\shortauthors{Kawasaki et al.}

\begin{document}

\title{
\ASCA\ Observations of the Supernova Remnant \ic:  
Thermal Structure and Detection of Overionized Plasma
}

\author{Masahiro Kawasaki\altaffilmark{1}, Masanobu Ozaki\altaffilmark{2},
         and Fumiaki Nagase\altaffilmark{3},}
\affil{Institute of Space and Astronautical Science, 
Sagamihara, Kanagawa, 229-8510, Japan}
\author{Kuniaki Masai\altaffilmark{4} and Manabu Ishida\altaffilmark{5}}
\affil{Department of Physics, Tokyo Metropolitan University, Hachioji, 
        Tokyo, 192-0397, Japan}
\and 
\author{Robert Petre\altaffilmark{6}}
\affil{Laboratory of High Energy Astrophysics, Code 662, 
        NASA/Goddard Space Flight Center, Greenbelt, MD 20771}

\altaffiltext{1}{kawasaki@astro.isas.ac.jp}
\altaffiltext{2}{ozaki@astro.isas.ac.jp}
\altaffiltext{3}{nagase@astro.isas.ac.jp}
\altaffiltext{4}{masai@phys.metro-u.ac.jp}
\altaffiltext{5}{ishida@phys.metro-u.ac.jp}
\altaffiltext{6}{rob@lheapop.gsfc.nasa.gov}

\begin{abstract}
We present the results of X-ray spatial and spectral studies of the
``mixed-morphology'' supernova remnant IC 443 using ASCA.  \ic\ has a
center-filled image in X-ray band, contrasting with the shell-like
appearance in radio and optical bands.  The overall X-ray emission is
thermal, not from a synchrotron nebula. ASCA observed IC 443 three
times, covering the whole remnant.  From the image analysis, we found
that the softness-ratio map reveals a shell-like structure.  At the
same time, its spectra require two (1.0~keV and 0.2~keV) plasma
components; the emission of the 0.2~keV plasma is stronger in the
region near the shell than the center.  These results can be explained
by a simple model that \ic\ has a hot (1.0~keV) interior surrounded by
a cool (0.2~keV) outer shell. From the emission measures, we infer
that the 0.2~keV plasma is denser than the 1.0~keV plasma, suggesting
pressure equilibrium between the two.  In addition, we found that the
ionization temperature of sulfur, obtained from H-like K$\alpha$ to
He-like K$\alpha$ intensity ratio, is 1.5~keV, significantly higher
than the gas temperature of 1.0~keV suggested from the continuum
spectrum.  The same can be concluded for silicon.  Neither an
additional, hotter plasma component nor a multi-temperature plasma
successfully accounts for this ratio, and we conclude that the 1.0~keV
plasma is overionized. This is the first time that overionized gas has
been detected in a SNR.  For the gas to become overionized in the
absence of a photoionizing flux, it must cool faster than the ions
recombine.  Thermal conduction from the 1.0~keV plasma to the 0.2~keV
one could cause the 1.0~keV plasma to become overionized, which is
plausible within an old (3$\times$10$^4$~yr) SNR.
\end{abstract}


\section{Introduction}\label{intro}
Supernova remnants (SNRs) have been traditionally classified into
three categories: shell-like, Crab-like (Plerionic; its emission
mainly originates from a pulsar nebula), and composite (shell-like
containing a Plerion). These SNRs usually have similar X-ray and radio
morphologies.  In addition to these, there is a recently established
class characterized by a center-filled morphology and thermal emission
in the X-ray band, contrasting with a shell-like structure in the radio and
optical bands.
\citet{rho98} termed these ``mixed-morphology'' SNRs (MM SNRs). The
distinctive characteristics of this category are: (i) the dominant X-ray
emission mechanism is thermal despite an X-ray morphology similar to
Crab-like SNRs; (ii) the emission arises primarily from the swept-up
interstellar material (ISM), not from ejecta; and (iii) they are
usually near or interacting with molecular clouds or other large ISM
structures.

\citet{rho98} identified two formation mechanisms that might explain the
properties of MM SNRs.  One is the cloud evaporation model, in which the
interior X-ray emission arises from the gas evaporated from
shocked clouds \citep{white91}. For this model to match the properties
of MM SNRs, the clouds must be numerous, small in size,
and of sufficient density to survive
in the passage of a strong shock. They pass into the hot interior
of the remnant, gradually evaporating and emitting X-rays.  The other is
the thermal conduction model \citep{cox99,shelton99b}.  
In this model, the shell of an expanding SNR cools below
$\sim$10$^6$~K and its X-ray emission becomes too soft to be detected
through the absorbing ISM\@. The thermal energy stored in the
shocked interior gas is conducted outward, producing the observed
uniform interior temperatures.  This model requires sufficient
interior density for conduction to be viable.  

\ic\ (G189.1+3.0) is one of the prototypical MM SNRs.  
It has a $\sim$45\arcmin\ diameter and is located at the distance of
1.5~kpc \citep{fesen84}.  It is strongly interacting with a foreground
molecular cloud and an \ion{H}{1} cloud to the east
\citep{cornett77,giovanelli79}.  It is a 
candidate for the CGRO EGRET $\gamma$-ray source 3EG J0617+2238 
\citep{hartman99}.  Its X-ray morphology was observed by
\Einstein\ to be center-filled, which is highly atypical of a SNR in
the adiabatic phase, with little correlation with the radio or optical
image. The spectrum of the brightest area
was well described by either a 1.5~keV non-equilibrium ionization
(NEI) plasma or 0.2~keV and 0.95~keV ionization equilibrium plasma
components \citep{petre88}.
\citet{asaoka94} analyzed \ROSAT\ all-sky survey and pointing data,
and found another partially overlapping SNR of $\sim$10$^5$~yr age,
which they called G189.6+3.3.  They reproduced the spectra for the
regions excluding the part overlapping with G189.6+3.3 by a
two-temperature (1~keV and $\sim$0.3~keV) ionization equilibrium plasma
model.

Using \Ginga, \citet{wang92} found a hard
X-ray component in the spectrum of \ic\ extending up to 20~keV.  
They suggested that this component originates from a very 
high temperature ($\gtrsim$10$^8$~K) plasma and thus the remnant's
age is about 10$^3$~yr instead of the previous estimate of 10$^4$ yr.
However, \ASCA\ and \sax\ revealed that this hard component is
non-thermal and originates in two small regions along the southern
edge of the radio shell where the SNR-molecular cloud interaction is
strongest \citep{keohane97,bocchino00}. Using \Chandra\ data, 
\citet{olbert01} determined that
one of the regions is a synchrotron nebula powered by a compact source
(probably a neutron star), whose kick velocity is consistent with an
age of 3$\times$10$^4$~yr.

In this paper, we first present the \ASCA\ GIS images and softness
ratio map (\S~\ref{ima}).  This reveals the existence of a soft,
shell-like structure.  We compare the SIS spectra of the north region
near the shell with that of the central region (\S~\ref{diff}) and
then perform the spectral fitting (\S~\ref{spe}).  Finally, we discuss
a possible mechanism that can account for our results on the
ionization state of \ic\ (\S~\ref{discuss}).

\section{Observations}\label{obs}
\begin{figure}
\epsscale{0.50}
\plotone{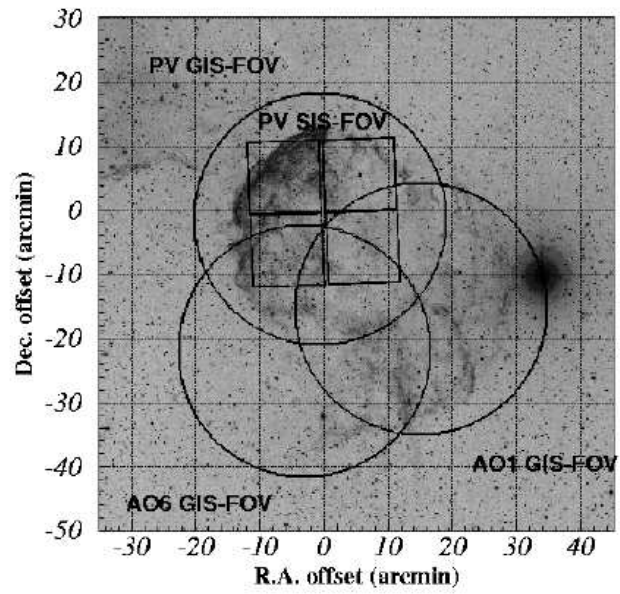}
\caption{Fields  of view of \ASCA\ GIS and SIS used in the
current analysis of \ic\ (20\arcmin\ radius region for GIS and whole
SIS field of view) superposed on the Palomar Observatory Sky Survey red image. 
Offsets indicated
are in arcminutes relative to the adopted center at R.A.=
06\h17\m20\s, Dec.=22\arcdeg40\arcmin50\arcsec\ (J2000).
\label{fig-fov}}
\end{figure}

\ASCA\ has two Solid-state Imaging Spectrometers (SIS0, SIS1) and Gas
Imaging Spectrometers (GIS2, GIS3) at the focal planes of the X-ray
telescopes (XRT). For \ic, three observations were carried out.
Figure~\ref{fig-fov} shows the fields of view (FOVs) of the
observations in the current analysis (20\arcmin\ radius region in GIS
and entire SIS FOV) overlaid on the Palomar Observatory Sky Survey red
image.  The relevant information about these observations is given in
Table~\ref{tbl-obs}. We extracted these data from the \ASCA\ public
archive.  We used all available GIS data.  During
the AO-1 phase, the GIS3 pulse height analyzer suffered from 
an on-board processing
problem: the least significant PHA bits were stuck, resulting in a
loss of spectral resolution.  Therefore the GIS3 AO-1 data used here
have only 8 binned PHA channels.
For the SIS, we used only the PV phase
observation data.  There are enough photons from this observation 
to allow investigation of
the spatial variation of the X-ray spectra with good energy
resolution.

For the most part, the NASA/GSFC revision~2
standard data processing criteria were used for screening all the GIS
and SIS data.  The lone exception was for the earth elevation angle
for the SIS:  in order to avoid contamination by stray light, we
used only  SIS0 data taken at angles $>$25\arcdeg\, and SIS1 data taken
at angles $>$20\arcdeg.
Since the PV phase
SIS data show no radiation damage effects, we did not apply the
Residual Dark Distribution (RDD) correction.

As \ic\ is located in the Galactic anti-center region ($l$,$b$) =
(189.1\arcdeg, 3.0\arcdeg) where the Galactic Ridge emission is
negligible, we used blank-sky data that include the cosmic X-ray
background (CXB) and the non-X-ray Background (NXB) for background
subtraction.  The PV phase GIS data, with no spread discriminator
application, contain more NXB events than those in subsequent observations.
For these data, we used the background datasets that are applicable to
observations performed when the GIS spread discriminator was turned
off.

In the present analysis, we used DISPLAY45 version 1.90 for the image
analysis, and XSPEC version 11.0 for the spectral analysis.

\section{Data Analysis and Results}\label{ana}

\subsection{Image Analysis}\label{ima}
\begin{figure}
\epsscale{0.32}
\plotone{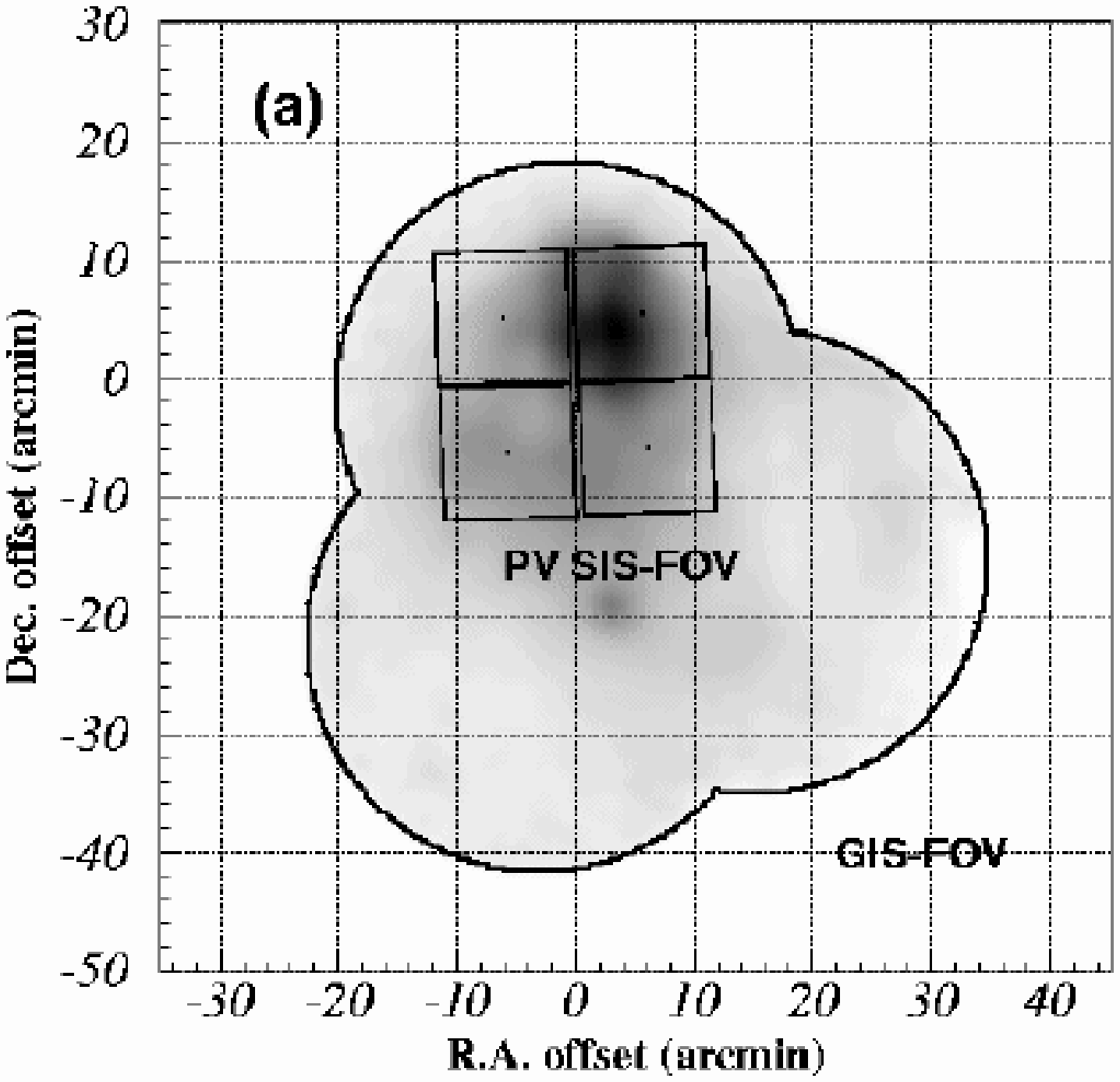}
\plotone{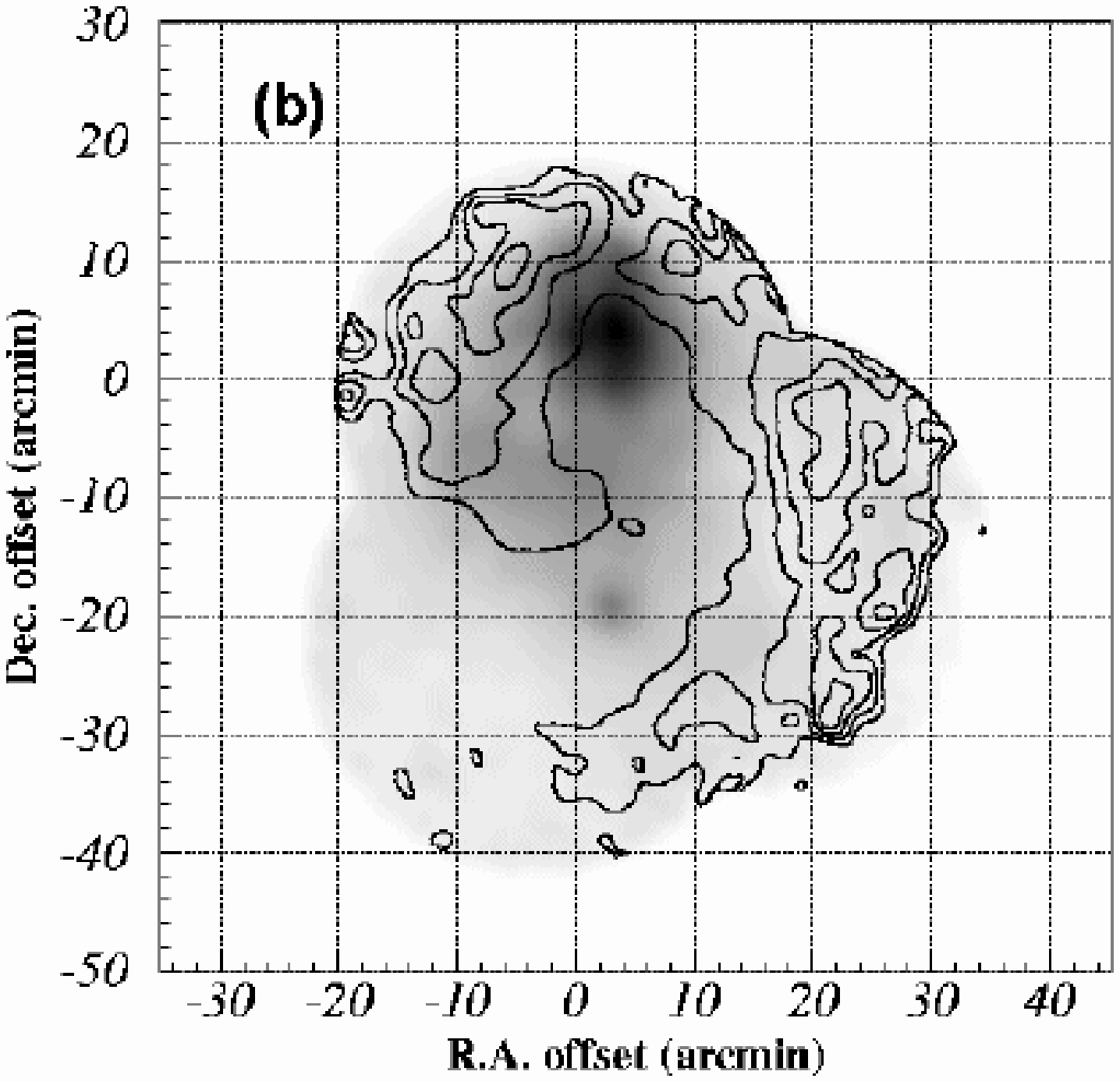}
\plotone{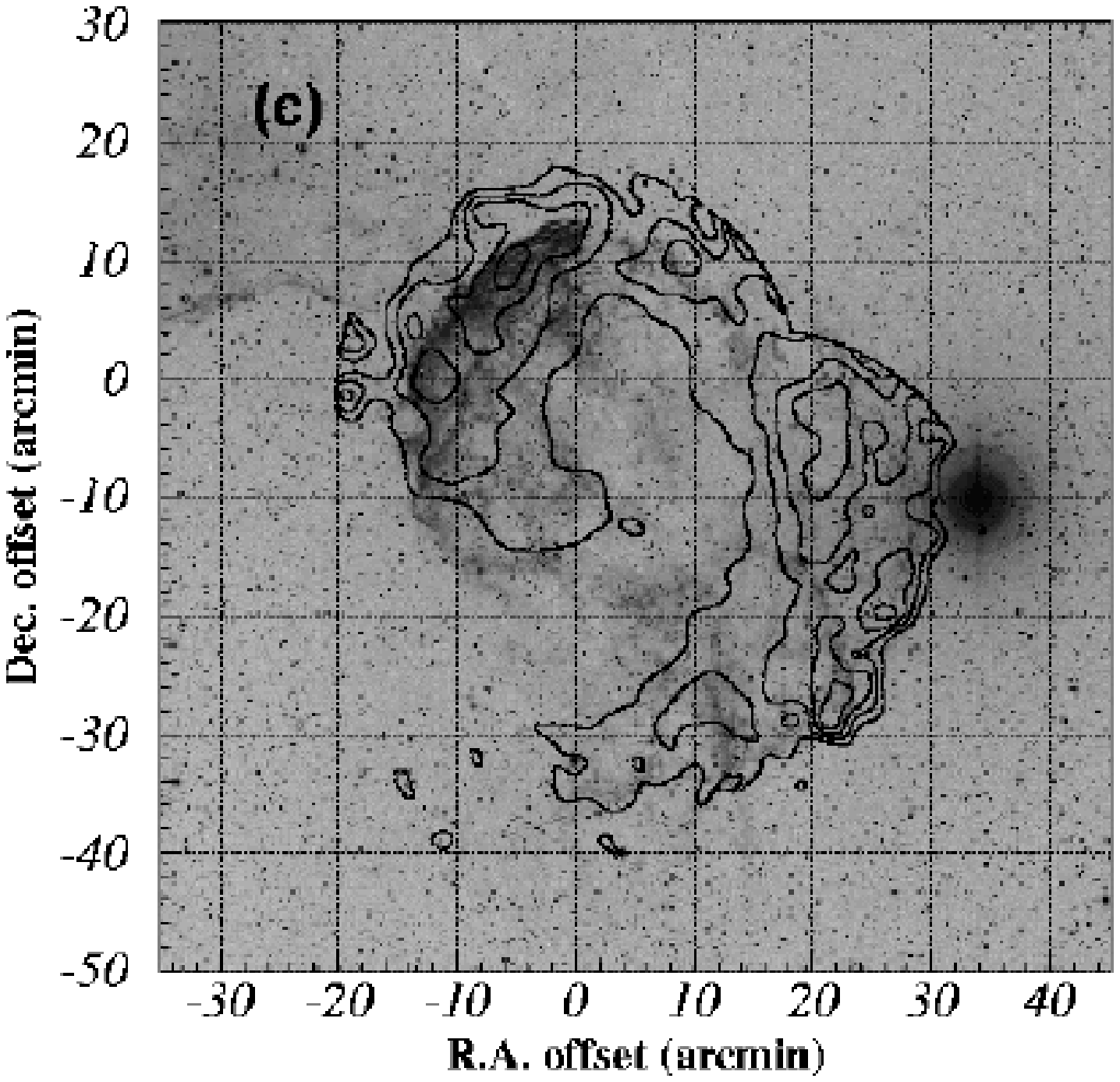}
\caption{(a): Vignetting- and exposure-corrected GIS 0.7--10.0~keV
band image with both SIS and GIS FOV boundaries.  GIS softness ratio
($F_{\mathrm{0.7-1.5~keV}}/F_{\mathrm{1.5-10.0~keV}}$) contours over
(b): GIS 0.7--10.0~keV band image, and (c): Palomar Digital Sky Survey
image.  The scale of softness-ratio contours is linear and their
levels are 60\%, 70\%, 80\%, 90\% of the peak value. Grayscales of the
GIS and sky survey images are linear and log respectively.  Offset
center of each figure is R.A.=06\h17\m20\s,
Dec.=22\arcdeg40\arcmin52\arcsec.  The contours are limited at the FOV
edge of the GIS along the northwest boundary, indicating the remnant
extends further.  The spot-like structure seen at (2,-20) in the GIS
image is the non-thermal component so-called HXF
\citep{keohane97,bocchino00,olbert01}.
\label{gis}
}
\end{figure}

We generated a complete image of \ic\ using GIS data.  The three GIS2
and GIS3 images were overlaid, and an exposure-corrected blank-sky
image was subtracted from it.  We multiplied the simulated XRT
efficiency map by the GIS grid map to form the vignetting image. Each
map is smoothed by the corresponding point spread function (PSF). We
constructed the exposure-corrected vignetting image, and divided it
into the source image to create a vignetting- and exposure-corrected
composite image. Figure~\ref{gis}(a) shows the resultant GIS
image in the 0.7--10.0~keV band with ASCA FOVs.

In every energy band, the composite image shows \ic\ to have a
centrally-filled morphology.  The remnant is larger in images below
1.5~keV than in higher energy bands. Thus a softness-ratio
($F_{\mathrm{0.7-1.5~keV}}/F_{\mathrm{1.5-10.0~keV}}$) map reveals a
shell-like structure, contrasting with the center-filled appearance in
the X-ray maps (see Fig.~\ref{gis}(b)).  As shown in Figure~\ref{gis}(c),
the soft shell corresponds well with the bright optical emission in
the northeast and southwest.  The contours are limited at the FOV edge along
the northwest boundary, indicating that it extends further.  The
absence of the softness-ratio shell in the southeast and its low
contrast in the northwest may be due to the absorption by the
molecular cloud in front of \ic\ \citep{cornett77}.

\begin{figure}
\epsscale{0.95}
\plottwo{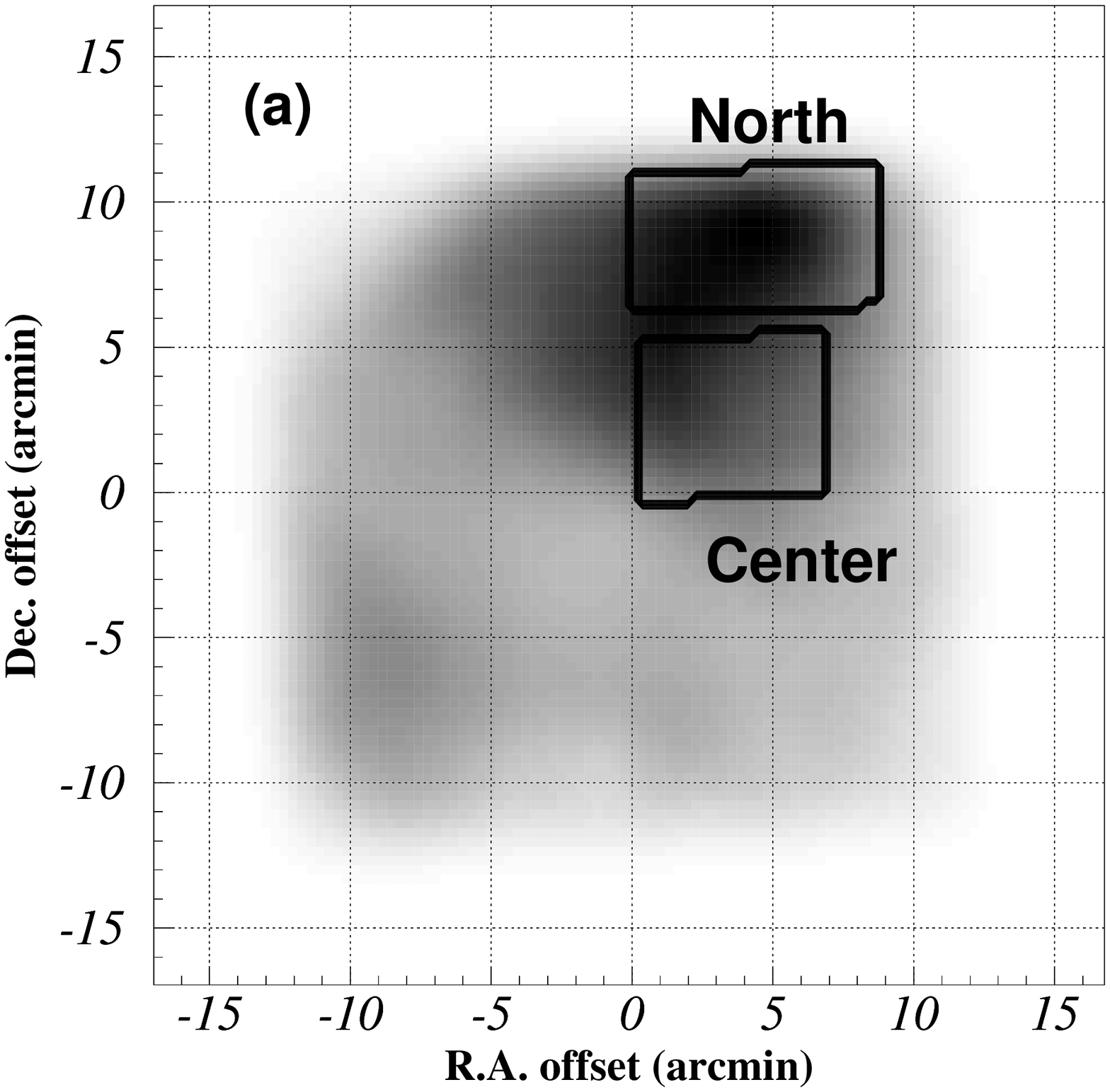}{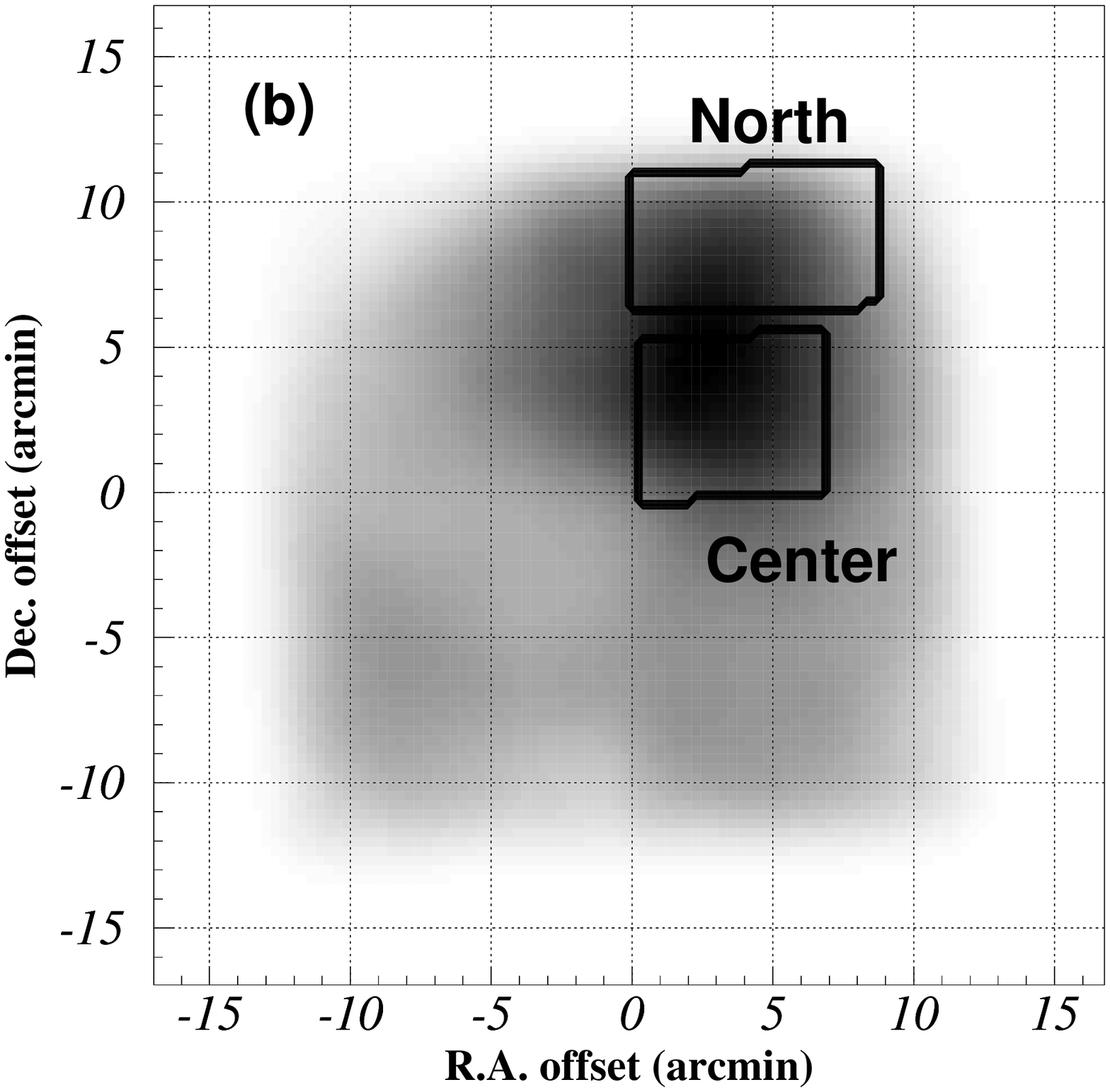}
\caption{Exposure- and vignetting-corrected SIS images of (a):
0.5--1.0~keV, and (b): 1.0--2.0~keV bands.  Two boxes are spectra
extraction regions named ``North'' and ``Center''.  Offset center of
each figure is R.A.=06\h17\m20\s, Dec.=22\arcdeg40\arcmin52\arcsec.
\label{sis}
}
\end{figure}

In Figure~\ref{sis} we show background-subtracted and exposure- and
vignetting-corrected SIS images in the 0.5--1.0~keV and
1.0--2.0~keV bands.  The SIS FOV includes both the X-ray
brightest region of \ic\ and a part of the northeast shell.
The peak surface brightness of the soft
(0.5--1.0~keV) component appears to the north of that of
the hard (1.0--2.0~keV) component (i.e. closer to the rim).

Therefore, both the GIS and SIS images suggest that a soft
($\lesssim$1.0~keV) X-ray component dominates near
the shell while a hard ($\gtrsim$1.5~keV) one dominates in the interior.

\subsection{Difference between Two Regions}\label{diff}

\begin{figure}
\epsscale{0.95}
\plottwo{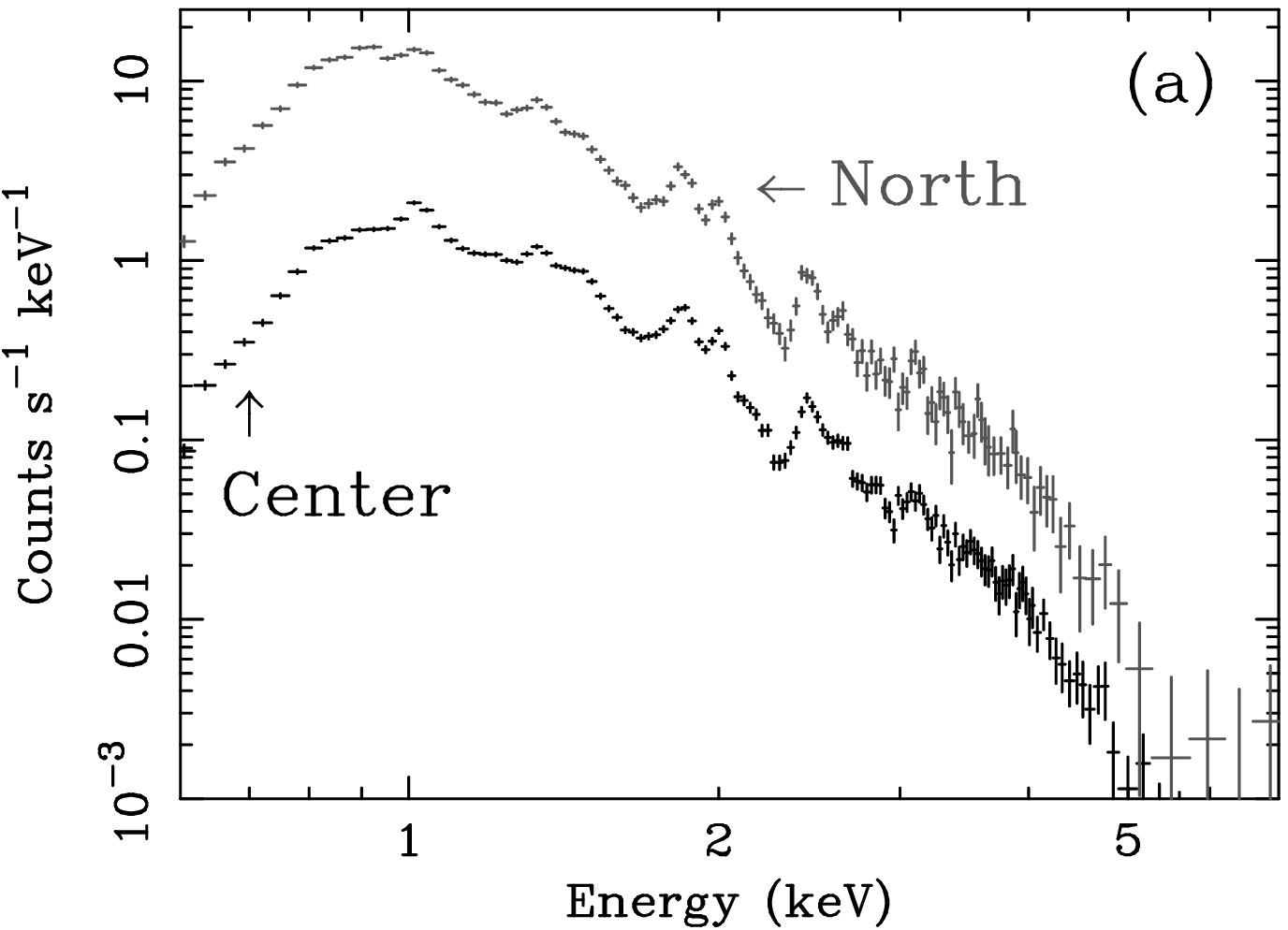}{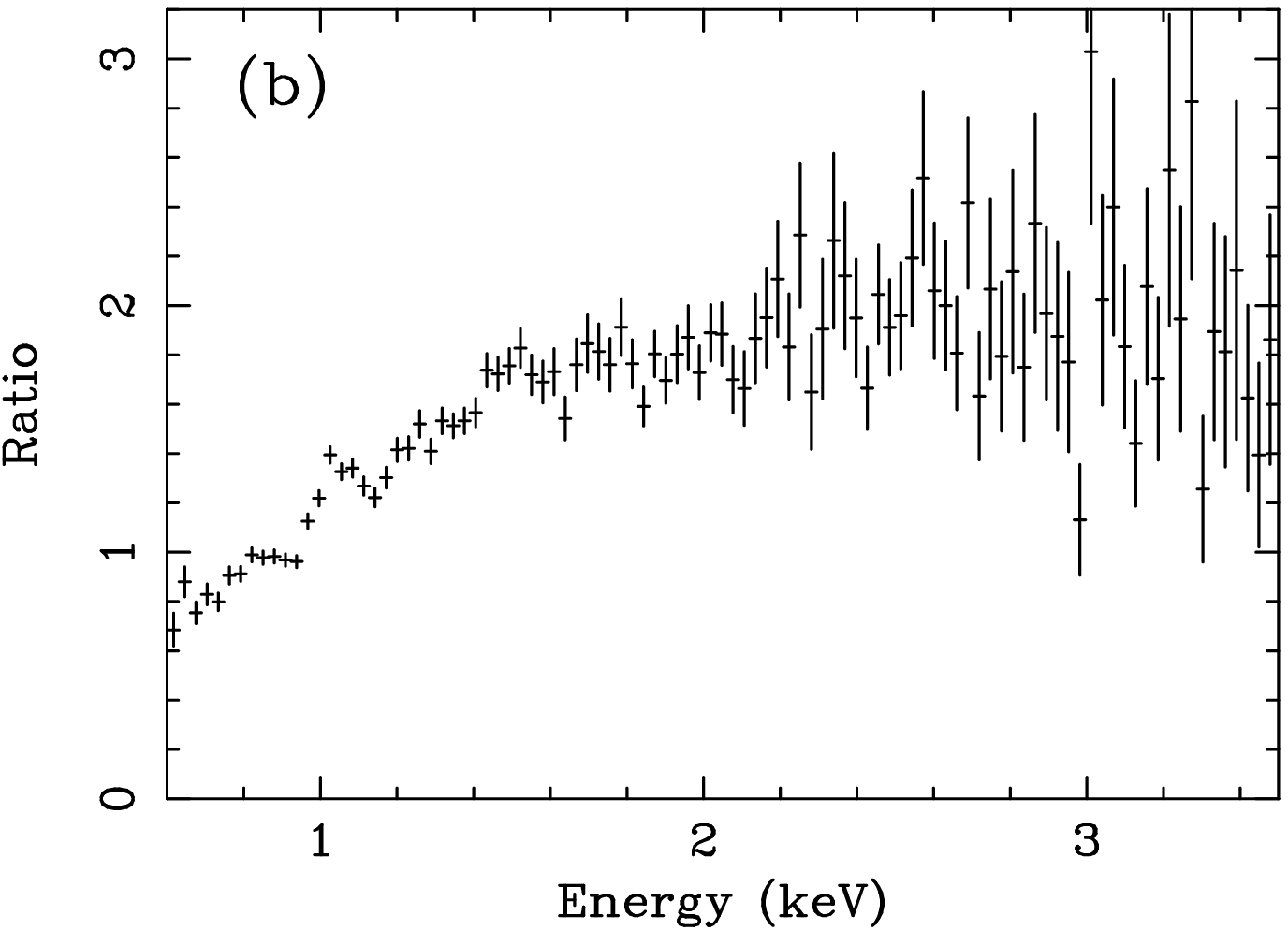}
\caption{SIS spectra of Center and North regions. 
(a): The left panel shows the Center and North spectra in black and gray
respectively. The spectrum of the North region is re-normalized to ten
times the original.  (b): The right panel shows the intensity ratio of
Center to North without any re-normalization. Below 1.4~keV, the ratio
decreases proportionally with energy.
\label{spe_diff}
}
\end{figure}

We therefore searched for a spectral variation from the interior to the
rim.  We used only SIS data for the spectral analysis because the SIS
has 
superior energy resolution to the GIS (FWHM of 5\% at 1.5~keV during the
observation \citep{tanaka94}) and a larger effective area below 1~keV.  
We extracted spectra from the two regions indicated by
boxes in Figure~\ref{sis}: a region near the northern rim (hereafter
North) where the soft X-ray emission is stronger, and an interior
region (hereafter Center) where the hard emission is stronger.
Neither region overlaps G189.6+3.3, and stray light from
the outside of the SIS FOV is negligible as shown in Figure~\ref{gis}(a).
The spectra and their ratio are shown in Figure~\ref{spe_diff}.

Above 1.4~keV, the shapes of the spectra appear similar to each other
(see Fig.~\ref{spe_diff}(b)), and are in good agreement with the
spectrum from a thin thermal plasma with a temperature of about 1~keV.
Below 1.4~keV, on the other hand, the North spectrum has relatively
stronger emission than the Center, and shows the \ion{Ne}{9} emission
line at 0.92~keV.  The intensity ratio of the Center spectrum to the
North below 1.4~keV is proportional to the energy, as shown in
Figure~\ref{spe_diff}(b). This ratio would be an exponential function
of energy if the difference were due to higher column density.
Therefore, it is difficult to explain this difference as a result of
column density variation.

\subsection{Spectral Analysis}\label{spe}
In order to characterize the spectral properties and evaluate the
difference between the two regions quantitatively, we performed model
fittings; first with a one-component plasma model, and then with a
two-component plasma model.  Since the SIS quantum
efficiency changes greatly at the energy of the oxygen edge
(0.53~keV), we did not use the data below 0.6~keV.

\subsubsection{One-Component Plasma Model}
\label{1plasma}
We first applied a simple single-temperature,
single-ionization-timescale, non-equilibrium ionization (NEI) plasma
model (\texttt{NEI} in XSPEC, \citet{borkowski00}) that was favored in
the previous \ic\ analysis \citep{petre88}.  However, the minimum
values of reduced $\chi^2$ were 924/118 for the North and 1337/128 for
the Center, even with variable abundances, which are hardly
acceptable.

We therefore tried an ad-hoc model consisting of thermal bremsstrahlung,
plus 12 Gaussian profiles to represent the most prominent emission
lines above 0.9~keV. The line widths and strengths were all left free.  Such
models have been used previously to model \ASCA\ spectra for other
remnants (e.g. \citet{miyata94,holt94}).  Both H-like and He-like
K$\alpha$ lines of Mg, Si, and S were well fitted with negligible line
widths compared with the energy resolution of the SIS.

\begin{figure}
\epsscale{0.50}
\plotone{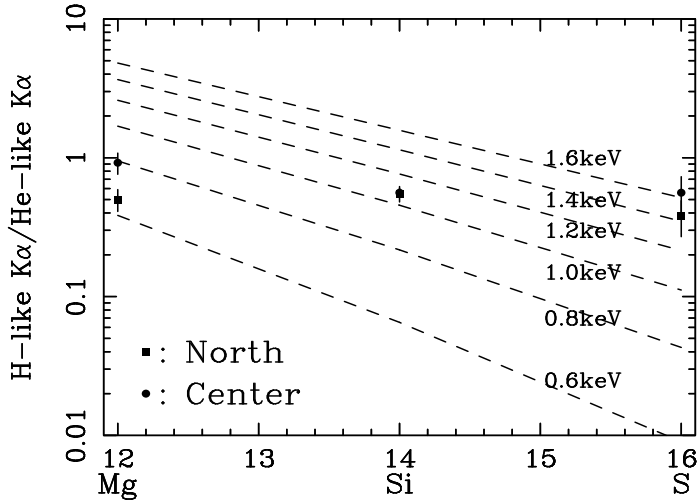}
\caption{Ratios of the fitted Gaussian fluxes of H-like K$\alpha$ to 
that of He-like K$\alpha$ for Mg, Si, and S.  Black squares and
circles are those data from the North and Center respectively.  This
ratio represents the ionization degree of each element. Dashed line
represents the ratios of H-like K$\alpha$ to 
that of He-like K$\alpha$ with each ionization temperature 
calculated based on \citet{mewe85}. All data are on different lines 
of ionization temperature.
\label{ion}
}
\end{figure}

The flux of each Gaussian profile is proportional to the
emissivity of the line component. The line intensity ratio of H-like
K$\alpha$ to He-like K$\alpha$ (hereafter \ratio) provides a measure
of the degree of ionization of a given element, and depends on both
the electron temperature and the ionization timescale.  
We evaluated the \ratio\ for Mg, Si, and S through comparison
with the theoretical values \citep{mewe85} as shown in
Figure~\ref{ion}, and found that it is impossible to explain the degree of
ionization of all elements with the same ionization temperature in
each region. This suggests that more than one plasma
component is required.

\subsubsection{Two-Component Plasma Model}
\label{2plasma}
We next applied a two-component plasma model.  The continuum components
above 1.4~keV of the two spectra have a shape consistent with a
temperature of 1~keV, as mentioned previously.  In
addition, Si and S appear to be ionized enough to apply a
collisional ionization equilibrium (CIE) model at this temperature
(see Fig.~\ref{ion}). We thus used the CIE plasma model
(\texttt{VRAYMOND} in XSPEC, \citet{raymond77}) as the high
temperature plasma.  On the other hand, we used the generalized NEI model
with varying temperature (\texttt{VGNEI} in XSPEC,
\citet{borkowski00}) as the low temperature plasma.  All 
abundances were treated as variables, and those of two
components linked together.
This model failed to reproduce the K$\alpha$ lines of \ion{Ne}{10},
\ion{Si}{14}, and \ion{S}{16} (and K$\beta$ line of \ion{Ne}{10} 
in Center), so we added several narrow Gaussian components.

\begin{figure}
\epsscale{0.95}
\plottwo{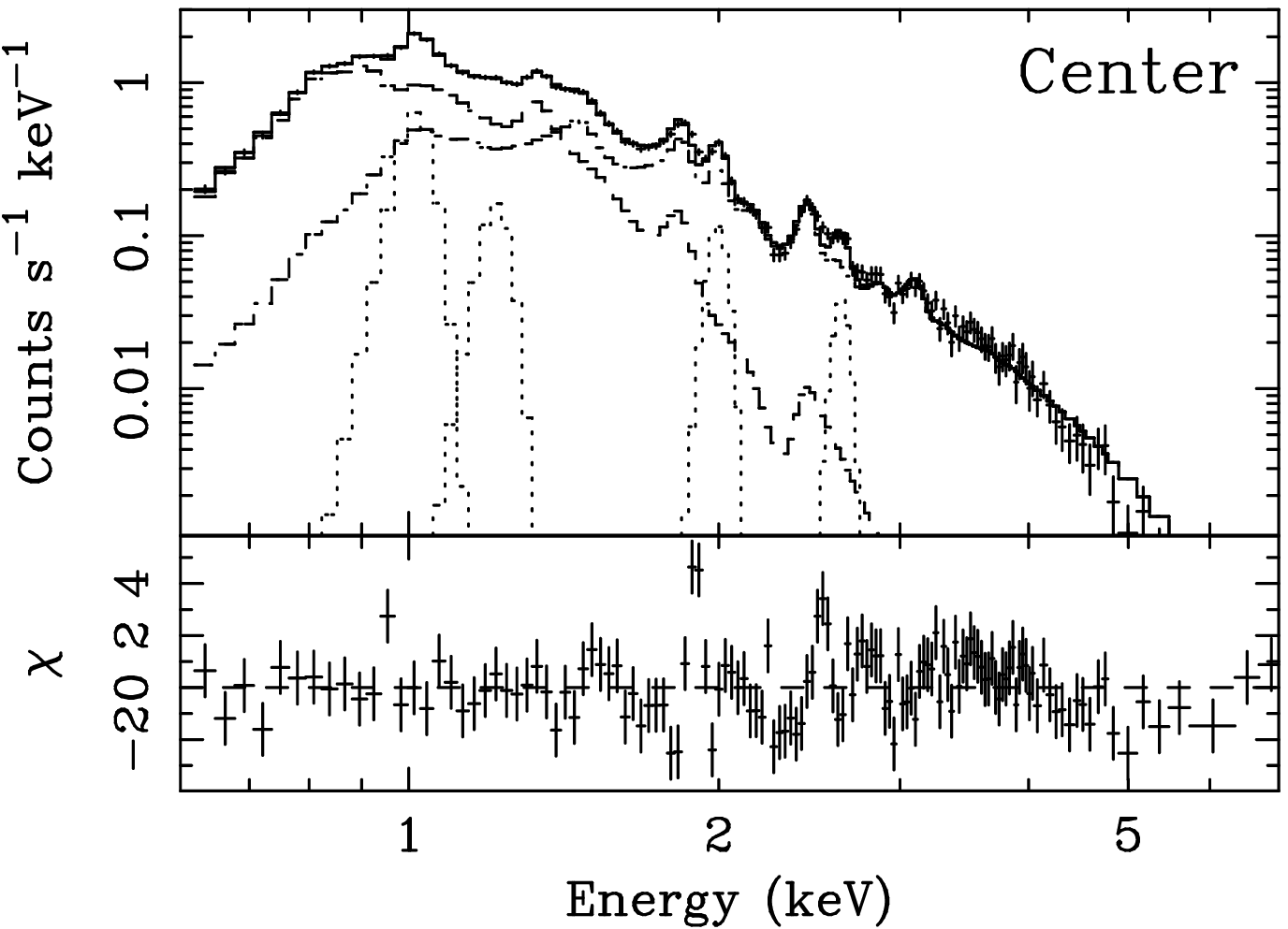}{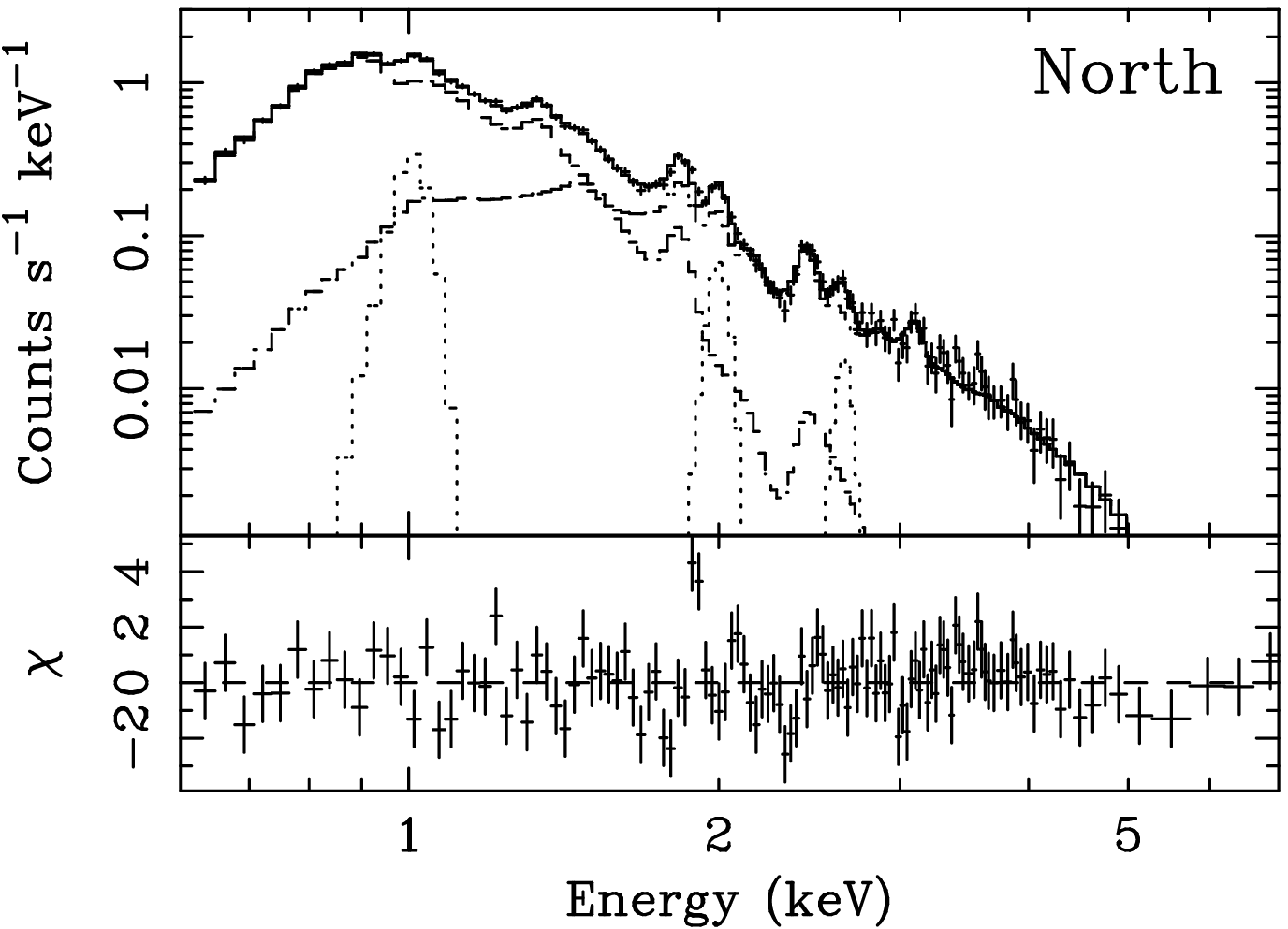}
\caption{ SIS spectra of Center (left panel) and North (right panel)
regions with best-fit two plasma and Gaussian models. Dashed,
dash-dotted, and dotted lines represent the 0.2~keV plasma, 1.0~keV
plasma, and additional Gaussians respectively.  The lower panels show
the residuals of the fit.
\label{spe-2pla}}
\end{figure}

We were able to fit reasonably both North and Center spectra
(except for the energies around the emission lines of Si and S) with a
model consisting of $\sim$0.2~keV NEI and $\sim$1.0~keV CIE plasma with
a column density of 7.4$\times$10$^{21}$~cm$^{-2}$ as shown in
Figure~\ref{spe-2pla}.  The best-fit parameters of the model are
listed in Table~\ref{tbl-2pla}.  All abundances are smaller than the
solar values \citep{anders89}, which suggests that the X-rays originate
from the shocked ISM.
Although these fits are not formally acceptable, the result implies
that the spectrum can be generally described by two plasma components:
Figure~\ref{spe-2pla} suggests that the spectra below 1.4~keV are well
reproduced by the low temperature ($\sim$0.2~keV) plasma,
whereas those above 1.4~keV show the features of the
1.0~keV plasma.

Almost all fitted parameters for the two spectra are consistent with each
other within the 90~\% confidence range. This suggests that the plasma
properties in the two regions are similar.  
The sole significant difference is the ratio of the
emission measures of the 0.2~keV plasma to the 1.0~keV one: it is
substantially higher in the North (185$^{+102}_{-68}$) than in the
Center (48$^{+39}_{-24}$).  Hence, we suggest that the difference of
the intensity ratio between two plasma components produces the
apparent difference of the spectra shown in
Figure~\ref{spe_diff}.

While the two component model accounts reasonably well for the gross
spectral features, it leaves an important detail unresolved.  The
H-like lines of Ne, Si and S are so strong that it is necessary to
add Gaussian components to represent them. 

\begin{figure}
\epsscale{0.95}
\plottwo{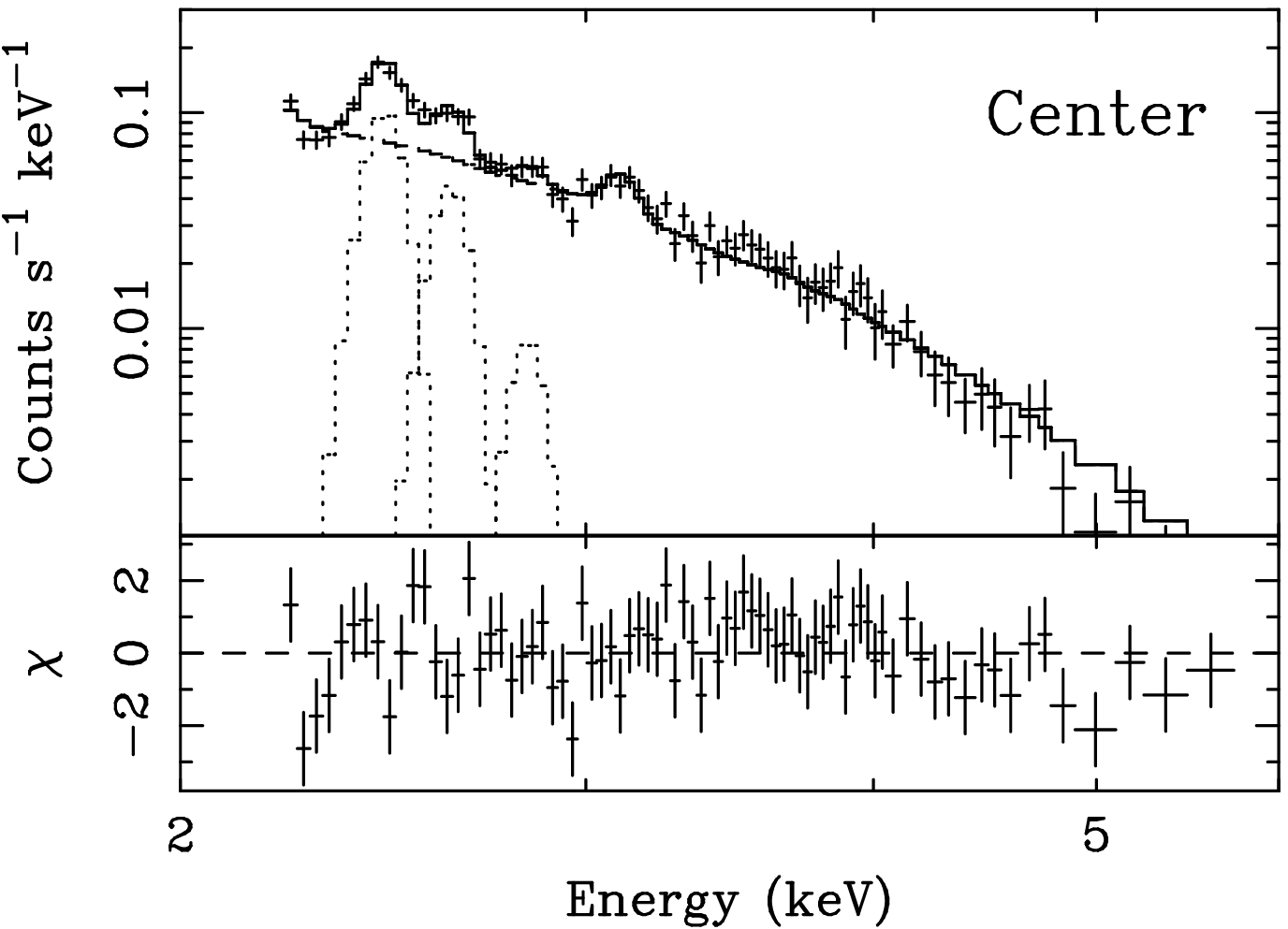}{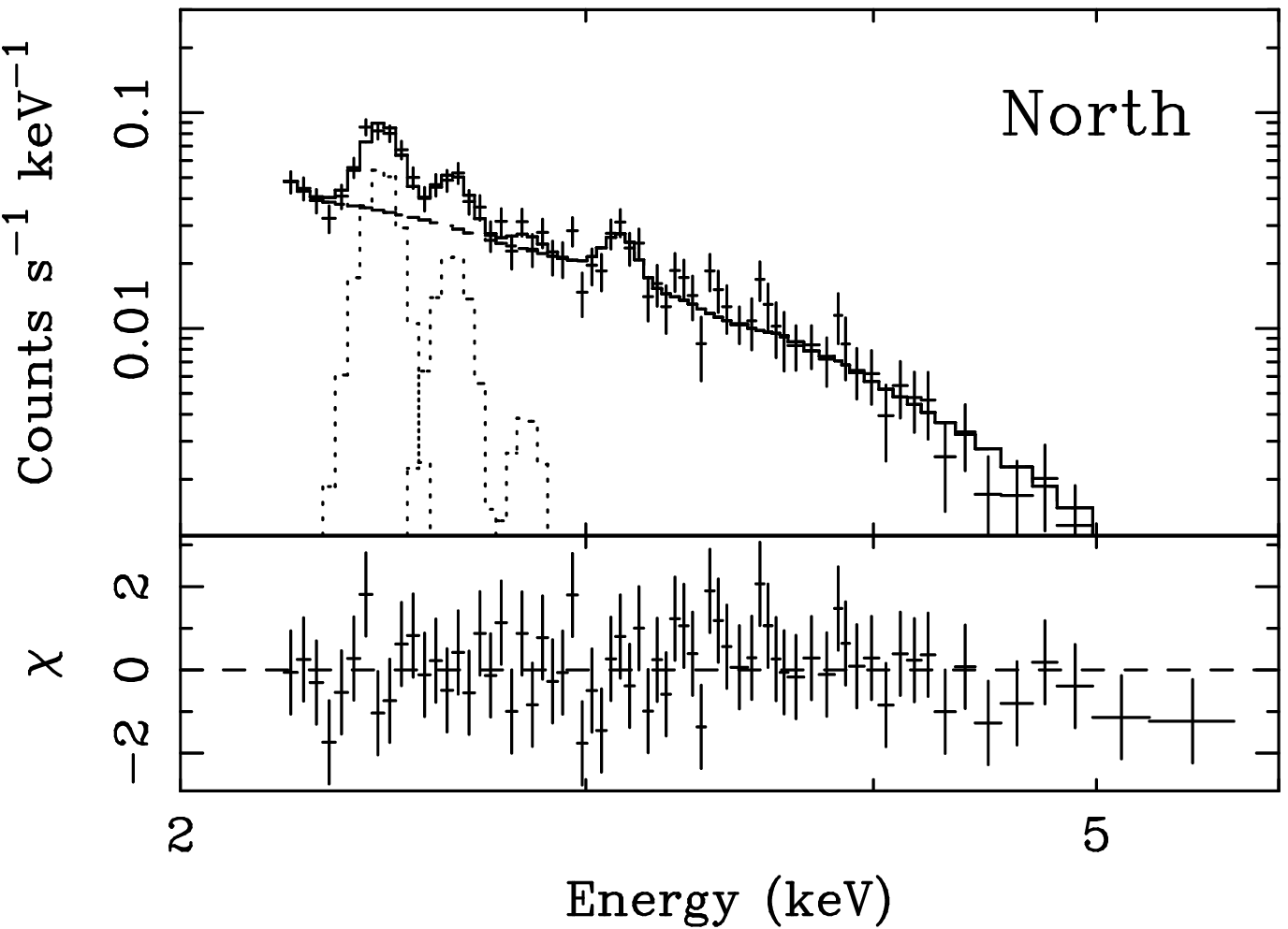}
\caption{ SIS spectra of Center (left panel) and North (right panel)
regions with the best-fit model of Raymond-Smith (dashed line) and
three narrow Gaussian components (dotted lines) with
$N_{\mathrm{H}}$=7.4$\times$10$^{21}$~cm$^{-2}$ at energies of
2.2--6.0~keV.  The lower panels show the residuals of the fit and the
minimum values of reduced $\chi^2$ are 85/70 and 53/59 in Center and
North respectively.
\label{spe-sulfur}}
\end{figure}

\begin{figure}
\epsscale{0.95}
\plottwo{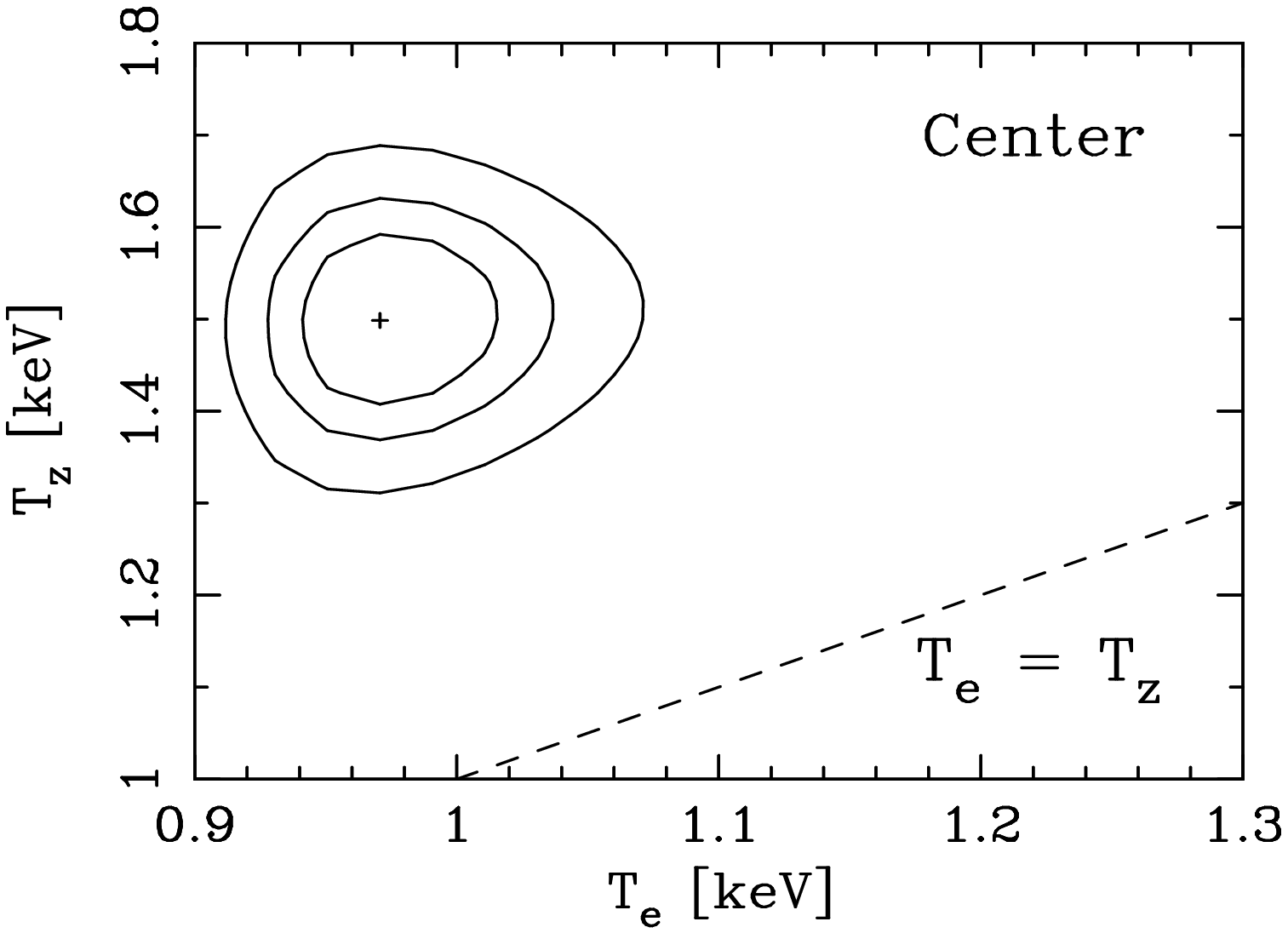}{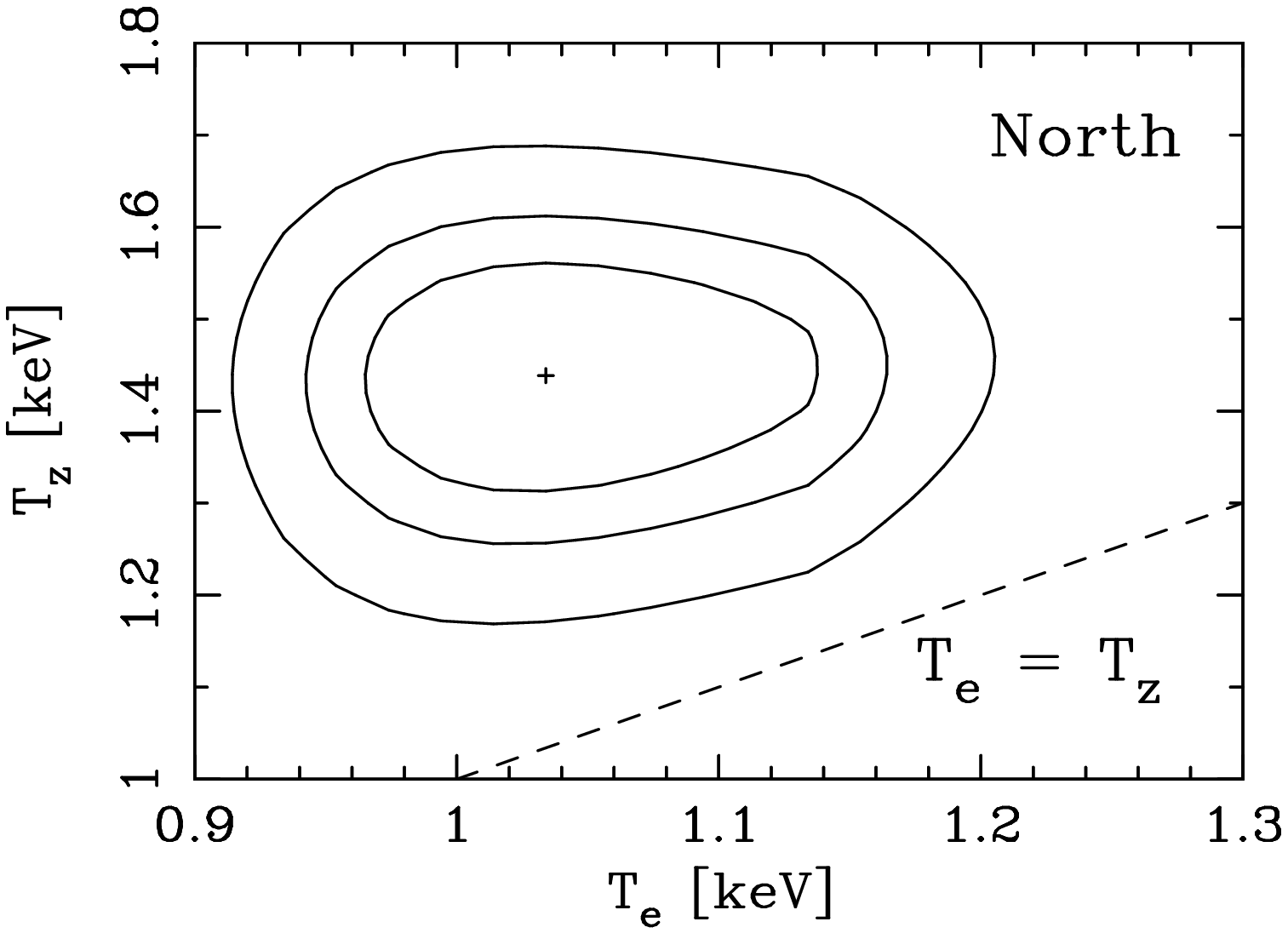}
\caption{Confidence contours of the ionization temperature
($T_z$) to the continuum temperature ($T_{\mathrm{e}}$) of
the SIS spectra in Center (left panel) and North (right panel).  The
confidence levels are 99\%, 90\%, and 67\%.  The contours are above the
dashed line which implies that the ionization temperature is higher
than the continuum temperature in each figure.
\label{cont}
}
\end{figure}

In order to investigate the anomalous S line ratio, we fit the 2.2--6.0~keV
spectra with a model consisting of a CIE plasma and three narrow Gaussian
components (\ion{S}{15} K$\alpha$, \ion{S}{16} K$\alpha$, \ion{S}{15}
K$\beta$). (We can ignore the contribution of the 0.2~keV plasma component at
the energies around the S lines.) 
The fits are acceptable as shown in Figure~\ref{spe-sulfur}.  The
\ratio\ for S requires an ionization temperature (hereafter $T_z$) of
about 1.5~keV in each region, higher than the continuum temperature
(hereafter $T_{\mathrm{e}}$) of 1.0~keV as shown in
Figure~\ref{cont}. The same can be concluded for Si.
\ion{Ne}{10} emits mostly 
around $kT_{\mathrm{e}}$ = 0.5~keV, which might indicate that the strong
\ion{Ne}{10} line comes from an intermediate temperature (0.2$<
kT_{\mathrm{e}}<$1.0~keV) plasma.

To account for the higher ionization temperatures of Si and S, we
replaced the Gaussians with a second, hotter (2.0~keV)
CIE component, whose temperature is selected to maximize the 
emissivity of the H-like line of S.  
However, this component requires an extraordinarily large
S abundance; more than 10$^3$ solar.  Such a large abundance is
not observed even in ejecta, and thus this model is physically
implausible.  Moreover, narrow band images at the energies
corresponding to Si and S H-like K$\alpha$ lines show no conspicuous
(e.g. clumpy) structure and are similar to those of other bands
(e.g. images for Si and S He-like K$\alpha$ and continuum bands),
indicating that the H-like lines are emitted from the same plasma that
emits He-like lines.  

Another possible source of the anomalous ratio between
$T_z$ and $T_{\mathrm{e}}$ is a temperature
distribution in the plasma, as observed in magnetic cataclysmic variables
\citep{ezuka99}. However, as in SNRs such plasma shows $T_z <
T_{\mathrm{e}}$, and thus does not explain the higher $T_z$ of Si and
S.  Therefore, we suggest that the K$\alpha$ lines of H-like Si and S
are emitted from an ``overionized'' 1.0~keV plasma (i.e., a plasma in
which the degree of ionization is larger than that expected from the
electron temperature assuming collisional ionization equilibrium).

\section{Discussion}\label{discuss}
\subsection{The Plasma Structure of \ic}
\ASCA\ observations have revealed the following new X-ray features of \ic; 
(1) the X-ray softness-ratio map shows a shell-like structure that
correlates with the optical filaments, implying
that the hard X-ray emission is more centrally concentrated than the
soft; (2) the spectra include two plasma components with temperatures
of 0.2~keV and 1.0~keV, and the intensity ratio of the former to the
latter is larger in the outer region than that in the inner region.

These results suggest that \ic\ has the plasma structure that can be
modeled by a central hot (1.0~keV) region, surrounded by a cooler
(0.2~keV) shell.  The mean electron densities within these regions are
estimated from the observed emission measures.  The emission measure
of the 1.0~keV plasma, which is estimated from the GIS spectrum of the
brightest region within 16\arcmin\ radius (north-east part of \ic\,
where most X-rays are emitted), is $\simeq1.1\times10^{58}
(d/1.5~\mathrm{kpc})^2$~cm$^{-3}$ ($d$ is the distance to \ic).  In
contrast, that of the 0.2~keV plasma is $\simeq5.4\times10^{59}
(d/1.5~\mathrm{kpc})^2$~cm$^{-3}$, more than an order of magnitude
larger.  To estimate the densities of the two components, we assume
that the 0.2~keV plasma extends to the north-east optical shell,
and thus has an outer radius of $\theta_2$ = 16\arcmin. The radius of
the region containing the
1.0~keV component is assumed to be $\theta_1$ = 10\arcmin,
considering that the surface brightness of the 1.0~keV at this
radius is 50 percent of its maximum
(from the 2.0--3.5~keV band image).  The densities
of the 1.0~keV plasma ($n_1$) and the 0.2~keV plasma ($n_2$) are
\begin{eqnarray}
n_1 &=& 1.0 \left( \frac{\theta_1}{10\arcmin} \right)^{-3/2} \left(
	\frac{d}{1.5~\mathrm{kpc}} \right)^{-1/2}~
\mathrm{cm^{-3}}, \label{n1}\\
n_2 &=& 4.2 \left( \frac{a^3-1}{3.096} \right)^{-1/2}
	\left( \frac{\theta_1}{10\arcmin} \right)^{-3/2} \left(
	\frac{d}{1.5~\mathrm{kpc}} \right)^{-1/2}~\mathrm{cm^{-3}},
\label{n2}
\end{eqnarray}
where $a = \theta_2 / \theta_1$.  The cooler exterior shell is denser
than the interior; the two components appear to be near pressure
equilibrium, and consistent with a typical SNR
predicted by the Sedov-Taylor solution, despite the fact that the X-ray 
image appears center-filled.

This structure is also consistent with the prediction of the
conduction model for mixed morphology SNRs, developed for W44
\citep{cox99,shelton99b}.  These authors  suggest that if a remnant, 
evolving in
a fairly smooth ambient medium of moderate density with thermal
conduction active in its interior, reaches the phase that radiative
cooling is important, then it shows inner-hot and outer-cool regions
of X-ray emission interior to the cold dense shell.  \ic\ has similar
characteristics to W44, including a partial \ion{H}{1} shell, suggesting it
is in the same evolutionary state.

\subsection{Formation Mechanisms of the Overionized Plasma}
Additionally, we found that the K$\alpha$ lines of H-like Si and S are
unusually strong, indicating that the plasma in \ic\ is overionized.
This is the first detection of evidence for an overionized plasma
in a SNR.  In SNRs, we are accustomed to consider the plasma
underionized, as in a collisionally-ionized, shock-heated plasma, the
ionization timescale is typically large compared with the time since
the bulk of the gas was shock heated.  Nevertheless, conditions can
(and apparently do) exist in which a shock-heated, low-density plasma
can become overionized.  In this section, we discuss possible
formation mechanisms for such a plasma.

We can consider two possible ways to produce an overionized plasma.
One is that photoionization causes heavy elements to be more highly
ionized than their collisional equilibrium level. A low-density plasma
does not produce sufficient radiation to photoionize the heavy
elements: the plasma needs a bright external radiation source.  CXOU
J061705.3+222127, with the 1--5 keV flux is $2 \times
10^{-13}$~erg~cm$^{-2}$~s$^{-1}$ \citep{olbert01,bocchino01}, is too
faint to photoionize the 1.0~keV plasma to emit the ``extra'' H-like
S flux of $\sim10^{-13}$~erg~cm$^{-2}$~s$^{-1}$ shown in
Table~\ref{tbl-2pla}, even if it was at the center of the remnant in
the past.
Futhermore, there is no other evidence of such a strong source in \ic\
and it is implausible to consider that this plasma region is
illuminated continuously by a hidden intense source.

The other way is that the gas cools faster than the ions can
recombine. The three possible cooling mechanisms in the gas are
radiation, expansion, and thermal conduction.
\citet{shelton98,shelton99a} has shown that inclusion of all three
processes in hydrodynamic modeling of SNRs expanding in a very low
density ISM leads to the development of an ``overionized'' plasma in
the interior once a remnant has reached the radiative stage.  The
\ion{H}{1} shell covering the eastern part of the remnant
\citep{giovanelli79} suggests that at least part of \ic\ is in the
radiative stage.  Below we compare the cooling timescales of radiation,
expansion, and conduction with the recombination timescale of heavy
elements in the 1.0~keV plasma, and show that these are
consistent with the formation of an overionized plasma in \ic.

The recombination timescale of an element of atomic number $Z$, which is
the characteristic timescale to reach ionization equilibrium, is
calculated by \citet{masai94}. In the case of sulfur at a temperature of 
1.0~keV, this is estimated to be
\begin{equation}
 t_{\mathrm{recomb}} \approx \sum_{z=0}^{Z} 
	\left( S_z + \alpha_z \right)^{-1} 
	\simeq 9 \times 10^{11} 
	\left( \frac{n_1}{1~\mathrm{cm^{-3}}} \right)^{-1}~\mathrm{s},
\label{trec}
\end{equation}
where $S_z$ and $\alpha_z$ represent the rate
coefficients for ionization and recombination from an ion of charge $z$ to
charge $z$+1 and $z-$1, respectively.  This is consistent with the
expectation that collisional ionization equilibrium is approached for
the ionization timescale value $n_{\mathrm{e}}t\sim10^{12}$~cm$^{-3}$~s.

We estimate the cooling timescale of the 1.0~keV plasma via
radiation to be
\begin{equation}
t_{\mathrm{cool}} = \frac{3 n_1 k T_1 V}{L_{\mathrm{rad}}} 
\simeq  4 \times 10^{14} 
\left( \frac{\theta_1}{10\arcmin} \right)^{3/2}
\left( \frac{d}{1.5~\mathrm{kpc}} \right)^{1/2} ~\mathrm{s},
\label{tcool}
\end{equation}
where $V$ and $L_{\mathrm{rad}}$ are the volume and the luminosity of the
1.0~keV plasma respectively.  The cooling timescale of
equation~(\ref{tcool}) is more than two orders of magnitude larger
than the recombination timescale of equation~(\ref{trec}). 

Rapid adiabatic cooling and the resultant overionization have been
discussed by \citet{itoh89} for a young remnant breaking out of the
circumstellar matter into a rarefied medium.  However, this is not the
case for \ic, which is presumed to be in its Sedov phase (expansion of
the shocked ejecta is negligible compared to the blast wave).  The
shock temperature decreases with blast wave expansion.  Even if the
effect is taken into account, it is found that the electron
temperature varies as slowly as $T_\mathrm{e} \propto t^{-2/25}$ with
Coulomb collisions in the post-shock region \citep{itoh78,masai94}.
Therefore, adiabatic expansion is unlikely to work, since the
timescale is much longer than the recombination timescale.


We next  calculate the energy transport rate by thermal conduction from the
1.0~keV interior to the 0.2~keV outer region.  We assume a temperature
gradient scale length, $l_{\mathrm{T}} \equiv (\mathrm{grad} \ln
T)^{-1}$, of (1--2)$\times$10$^{19}$~cm, the distance from the
approximate boundary between the 1.0~keV and the 0.2~keV components to the edge
of the remnant.  This distance is much longer than the mean free path of
electrons $\lambda_{\mathrm{e}} \approx 10^{18} \left( T_{\mathrm{e}}
/ 10^7~\mathrm{K}\right)^2 \left( n_{\mathrm{e}} / 1~\mathrm{cm^{-3}}
\right)^{-1} $ cm.  The total flow of heat to the outer plasma
is estimated to be
$2\times10^{38}(l_{\mathrm{T}}/10^{19}~\mathrm{cm})^{-1}(\theta_1/10\arcmin)^2
(d/1.5~\mathrm{kpc})^2$~erg~s$^{-1}$.  The energy
loss via radiation in the 0.2~keV plasma is
7$\times$10$^{37}$~erg~s$^{-1}$ from the observed emission measure,
comparable to the heat flow.  In this circumstance, thermal conduction
equilibrates the temperatures of the two components on 
the conduction timescale,
\begin{eqnarray}
t_{\mathrm{cond}} &\equiv& - 
	\left( \frac{d \ln T_{\mathrm{e}}}{dt} \right)^{-1}
	\approx \frac{n_{\mathrm{e}} l_{\mathrm{T}}^2 k}{\kappa} \nonumber\\
&\simeq& 2 \times 10^{11}
	\left( \frac{n_1}{1~\mathrm{cm^{-3}}} \right) 
	\left( \frac{l_{\mathrm{T}}}{10^{19}~\mathrm{cm}}\right)^2
	\left( \frac{kT_1}{1.0~\mathrm{keV}} \right)^{-5/2}
	\left( \frac{\ln \Lambda}{32.2} \right)~\mathrm{s},
\label{tcond}
\end{eqnarray}
where $\ln \Lambda$ is the Coulomb logarithm.  The gas in the 1.0~keV
plasma cools on the timescale given by equation~(\ref{tcond}). This
timescale is smaller than the recombination timescale, and is 
comparable to it even with $l_{\mathrm{T}} = 2 \times 10^{19}$~cm.  This
indicates that the strong gas cooling via conduction makes
$T_{\mathrm{e}}$ lower than $T_z$.  While the presence of a magnetic
field will in principle reduce the heat flux, we ignore its effect,
since the magnetic field is expected to be negligibly small in mature
SNRs.  We conclude from this argument that it is
entirely possible in \ic\ and similar SNRs for the interior gas to be
an overionized, recombining plasma.

Assuming both $T_z$ and $T_{\mathrm{e}}$ are $e$-folding with their
characteristic timescales, we can estimate the age of the 1.0~keV
plasma ($t_{\mathrm{age}}$) approximately by the following equation,
\begin{equation}
\frac{T_z}{T_{\mathrm{e}}} \approx \frac{\exp \left[
	- t_{\mathrm{age}}/ t_{\mathrm{recomb}}\right]} {\exp \left[ -
	t_{\mathrm{age}}/{t_{\mathrm{cond}}} \right]} \sim 1.5,
\label{tratio}
\end{equation}
to be (0.3--1)$\times$10$^4$~yr. This age is younger than that estimated by
\citet{chevalier99}. It indicates at least that the 
1.0 keV plasma is younger than $t_{\mathrm{recomb}} \approx 3
\times 10^4$~yr; if this plasma is older than the age represented by 
the recombination timescale, we would expect $T_z$ to be comparable to
$T_{\mathrm{e}}$.

Evidence of overionization has not been detected in any other SNR.  We
consider our detection in \ic\ to be plausible for the following reasons: 
(1) The interior of \ic\ was shocked long ago and has had no
subsequent source of heating or ionization, and thus should be much nearer
ionization equilibrium than newly shocked gas near the rim or in young
SNR's; (2) \ic\ has an
interior plasma hot enough to give rise to strong thermal conduction; (3)
\ASCA\ SIS is the first detector with sufficient energy resolution to
cleanly resolve the  H-like and He-like lines of S from each other; 
(4) The signal-to-noise in the spectra is sufficiently high 
((4--5)$\times10^4$ counts in each region) for
$T_z$ and $T_{\mathrm{e}}$ to be inferred with small errors.  
It should be noted,
however, that no evidence of overionization is detected in W44,
despite it having similar properties and comparable quality of \ASCA\ spectra.
The difference between the two remnants merits further investigation.

\section{Conclusion}\label{conc}
We have discovered  a shell-like structure in the softness-ratio map and two
distinct (0.2~keV and 1.0~keV) plasma components in \ic.  Using these
results and the emission measures of the two components, we suggest that
\ic\ has a plasma structure consisting of a central hot region
surrounded by an envelope of
a cool and denser plasma, that is, the typical structure of a SNR
predicted by the Sedov-Tayler solution.  In addition, we have detected
strong line emission from H-like Ne, Si and S.  The line intensity
ratio of H-like K$\alpha$ to He-like K$\alpha$ of S requires a
temperature of 1.5~keV, which is significantly higher than the
continuum temperature of 1.0~keV.  The same can be concluded for Si.
Therefore, we suggest that the 1.0~keV plasma is ``overionized.''

In order to produce an overionized thermal plasma, the gas cooling
rate must be higher than the recombination rate of ions.  Strong
cooling due to thermal conduction from the hot (1.0~keV) remnant
center to the cooler (0.2~keV) exterior could cause this.

\ic\ is the first SNR in which evidence for overionization has been found.  
The process of thermal conduction should arise in all SNRs. Therefore
we expect that observations of other evolved SNRs (such as
mixed-morphology types) should detect further evidence of overionized
plasma.  The apparent absence of overionized gas in the similar
remnant W44 suggests that the interior plasma in other remnants may
evolve differently from that in \ic.

\acknowledgments
We thank Yasunobu Uchiyama for his thoughtful comments and
suggestions.  The Digitized Sky Survey was produced at the Space
Telescope Science Institute under U.S. Government grant NAG
W-2166. The images of these surveys are based on photographic data
obtained using the Oschin Schmidt Telescope on Palomar Mountain and
the UK Schmidt Telescope. The plates were processed into the present
compressed form with the permission of these institutions.


\clearpage

\clearpage

\begin{deluxetable}{ccccc}
\tablecaption{\ASCA\ observation logs of \ic \label{tbl-obs}} 
\tablehead{
\colhead{} & \colhead{} &  \multicolumn{3}{c}{Observation Phase}\\ 
\cline{3-5} \\ 
\colhead{Parameter} & \colhead{} 
& \colhead{PV} & \colhead{AO-1} & \colhead{AO-6}
}
\startdata 
Sequence Number & & 10011010, 10011020 & 51023000 & 56057000 \\
R.A. (J2000) & & 06\h17\m26\s & 06\h17\m18\s & 06\h17\m34\s \\
Dec. & & 22\arcdeg40\arcmin48\arcsec & 22\arcdeg26\arcmin45\arcsec 
	& 22\arcdeg20\arcmin22\arcsec \\
Date & & 1993 Apr 14--15 & 1994 Mar 9--10 & 1998 Mar 25--27 \\
\hline
observation mode & & & & \\ \cline{1-2}
SIS & & 4CCD & 4CCD & 2CCD \\ 
GIS & & standard-PH & standard-PH & standard-PH \\ \hline
effective exposure-time & & & & \\ \cline{1-2}
SIS0/SIS1 & & 35.4/36.6 ksec & - & - \\
GIS2/GIS3 & & 39.7/39.7 ksec& 35.4/35.4 ksec & 37.5/37.5 ksec \\

\enddata
\end{deluxetable}

\begin{deluxetable}{ccccccc}
\tablecaption{
Best-fit parameters of the two plasma model \label{tbl-2pla}} 
\tablehead{
\colhead{} & \colhead{} & \colhead{} & 
\multicolumn{2}{c}{North Region} & \multicolumn{2}{c}{Center Region} \\ 
\cline{4-7} 
\colhead{model} & \colhead{parameter(unit)} & \colhead{} & 
\colhead{value} & \colhead{(90\% c.r.)} & 
\colhead{value} & \colhead{(90\% c.r.)}
}
\startdata
wabs & $N_{\mathrm{H}}$(10$^{21}$cm$^{-2}$) & & 
7.3 & (6.7--7.9) & 7.5 & (6.9--8.4) 
\\ \hline
 & $kT$(keV) & & 0.17 & (0.15--0.19) & 0.18 & (0.16--0.21) 
\\ 
vgnei & $n_{\mathrm{e}} t$(10$^{11}$ cm$^{-3}$~s) & & 
1.0 & (0.7--1.5) & 1.0 & (0.7--1.5) 
\\
(low $T$) & $<kT>$(keV) & & 0.44 & (0.39--0.53) & 0.51 & (0.46--0.59) 
\\
 & constant factor\tablenotemark{a} & & 
185 & (117--287) & 48 & (24--87)
\\ \hline
vraymond & $kT$(keV) & & 1.09 & (1.01--1.19) & 1.06 & (1.01--1.11)
\\
(high $T$) & emission measure\tablenotemark{b} & & 0.038 &
(0.030--0.046) & 0.056 & (0.048--0.064)
\\ \hline
 & O & & 0.03 & (0.02--0.04) & 0.07 & (0.05--0.10) 
\\
 & Ne & & 0.09 & (0.08--0.11) & 0.12 & (0.10--0.16) 
\\
 & Mg & & 0.11 & (0.09--0.14) & 0.25 & (0.19--0.35) 
\\
Abundance & Si & & 0.42 & (0.34--0.52) & 0.38 & (0.32--0.47) 
\\
 & S & & 0.43 & (0.35--0.54) & 0.43 & (0.38--0.49) 
\\
 & Ar & & 0.64 & (0.39--0.89) & 0.57 & (0.41--0.72) 
\\
 & Fe & & 0.07 & (0.05--0.09) & 0.17 & (0.14--0.20) 
\\ \hline \hline
\ion{Ne}{10} K$\alpha$ & observed energy (keV) & & 
	1.012 & (1.006--1.018) & 1.021 & (1.018--1.025) \\
	& flux\tablenotemark{c} & & 21 & (18--26) & 29 & (23--39) \\
\ion{Ne}{10} K$\beta$ & observed energy (keV) & & 
	- & - & 1.223 & (1.215--1.231) \\
	& flux\tablenotemark{c} & & - & - & 3.1 & (2.3--4.1) \\
\ion{Si}{14} K$\alpha$ & observed energy (keV) & & 
	2.015 & (2.002--2.011) & 2.011 & (2.002--2.018) \\
	& flux\tablenotemark{c} & & 1.1 & (0.8--1.4) & 1.3 & (1.1--1.6) \\
\ion{S}{16} K$\alpha$ & observed energy (keV) & & 
	2.664 & (2.641--2.679) & 2.649 & (2.630--2.665) \\
	&flux\tablenotemark{c} & & 0.31 & (0.20--0.42) & 0.53 & (0.43--0.64) 
\\ \hline 
\multicolumn{2}{c}{$\chi^2$/d.o.f} & & 160/108 & & 226/116 & 
\\
\enddata
\tablenotetext{a}{This constant factor present the ratio of the 
\texttt{VGNEI} emission measure to the \texttt{VRAYMOND} one in XSPEC.}
\tablenotetext{b}{The unit is 4$\pi d^2 \times 10^{14}$~cm$^{-5}$ where $d$ 
is a distance to \ic.}
\tablenotetext{c}{The unit is 10$^{-4}$ photons~cm$^{-2}$~s$^{-1}$.}

\tablecomments{ Each abundance is relative to the solar value and those of 
two plasma components linked together.\\ 
The 90\% confidence range (c.r.) for temperature, ionization
timescale, abundances, and line parameters are given in brackets.\\
The line width of each Gaussian profile is fixed to 0.  }
\end{deluxetable}

\end{document}